\documentstyle[sprocl]{article}

\input epsf
\def\be{\begin{equation}}
\def\ee{\end{equation}}
\def\bea{\begin{eqnarray}}
\def\eea{\end{eqnarray}}

\newcommand\lesssim{\mathrel{\rlap{\lower4pt\hbox{\hskip1pt$\sim$}}
    \raise1pt\hbox{$<$}}}
\newcommand\gtrsim{\mathrel{\rlap{\lower4pt\hbox{\hskip1pt$\sim$}}
    \raise1pt\hbox{$>$}}}

\begin{document}

\title{PRIMORDIAL BLACK HOLES AND EARLY COSMOLOGY}

\author{ANDREW R.~LIDDLE and ANNE M.~GREEN }

\address{Astronomy Centre, University of Sussex,\\
Brighton BN1 9QH, Great Britain}


\maketitle\abstracts{
We describe the changes to the standard primordial black hole 
constraints on density perturbations if there are 
modifications to the standard cosmology between the time of formation and 
nucleosynthesis.}
  
\section{Introduction}

Primordial black holes (PBHs) provide an important constraint on the physics 
of the 
very early Universe. They may form with masses low enough for Hawking 
evaporation to be important, which gives them a lifetime
\begin{equation}
\frac{\tau}{10^{17} \, {\rm sec}} \simeq \left( \frac{M}{10^{15} \, {\rm
	grams}} \right)^3 \,.
\end{equation}
From this we learn that a PBH of initial mass $M \sim 10^{15}$g will  
evaporate at the present epoch, while another interesting mass is 
$M \sim 10^{9}$g which leads to evaporation around nucleosynthesis. 
Evaporations of PBHs at these times are strongly constrained.

Several mechanisms have been proposed which might lead to PBH formation; the 
simplest is collapse from large-amplitude, short-wavelength density 
perturbations. They will form with approximately the horizon mass
\begin{equation}
M_{{\rm HOR}} \simeq 10^{18} \, {\rm g} \, \left( 
	\frac{10^7 \, {\rm GeV}}{T} \right)^2 \,,
\end{equation}
which tells us that the PBHs for which evaporation is important must 
have formed during very early stages of the Universe's evolution. The reason 
why the constraints are typically so strong is that after formation black 
holes redshift away as non-relativistic matter. In the standard cosmology the 
Universe is radiation dominated at these times, and so the energy density in 
black holes grows, relative to the total, proportionally to the scale factor 
$a$.

The purpose of this article is to emphasize that the constraints obtained 
depend not just on the formation rate and time of the black holes, but on the 
complete 
cosmological history. It has recently been fashionable to consider 
alternatives to the standard cosmology, especially in the interval between 
about $10^7$ GeV and nucleosynthesis. We have studied two such possibilities, 
a Universe with thermal inflation\cite{ls} and one with a prolonged period of 
matter 
domination at early times, as well as re-examining the standard cosmology.
Here we summarize the results, which have been reported in full in two 
papers.\cite{gl,glr}

The three different cosmological histories we study are as follows. In the 
standard cosmology we begin with inflation (which generates the density 
perturbations), which gives way through pre/reheating to radiation domination 
and eventually matter domination. The thermal inflation cosmology introduces 
a short period of inflation during the radiation-dominated era. Finally, the 
moduli-dominated scenario inserts a period of moduli domination, the moduli 
behaving like non-relativistic matter, during the radiation era.

\begin{figure}[t]
\centering 
\leavevmode\epsfysize=7.7cm \epsfbox{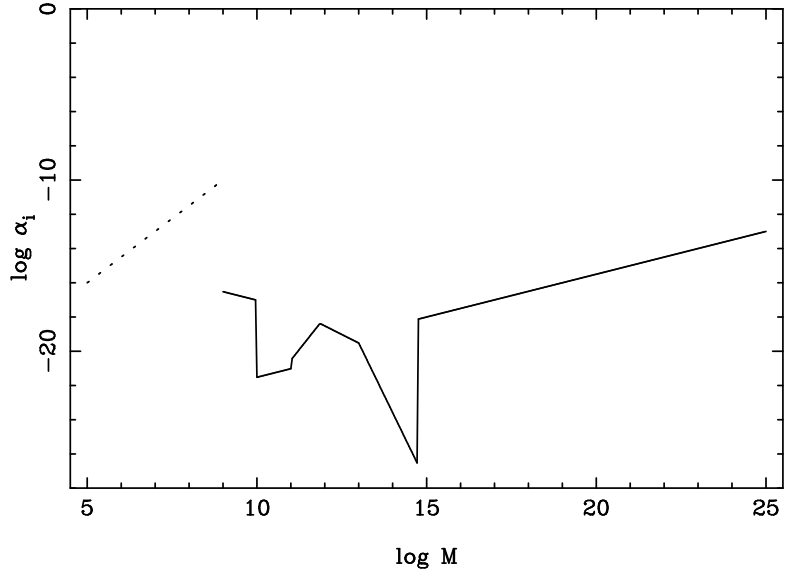}\\ 
\caption[fig1]{\label{fig1} Here $\alpha$ is the fraction of black holes 
permitted to form. The dotted line assumes black hole evaporation leaves a 
relic, and is optional.} 
\end{figure} 

\section{The standard cosmology}

The many limits \cite{limits,gl} on the black hole abundance in different 
mass ranges are shown in Fig.~\ref{fig1}. Typically no more than about 
$10^{-20}$ 
of the mass of the Universe can go into PBHs. This limits the size of density 
perturbations on the relevant mass scale. The minimum mass which can form is 
governed by the reheat temperature after inflation.

One way of using these constraints is to consider power spectra normalized to 
the COBE observations, which probe $M \sim 10^{56}$g.
The simplest example is to consider power-law spectra with a constant 
spectral index $n$ across all relevant scales. (Note that some hybrid 
inflation models predict $n$ constant even to very short scales.) 

The interesting situation is $n>1$, for which the shortest-scale 
perturbations dominate, and this was explored by Carr et al.~\cite{cgl} We 
have redone their analysis and corrected two significant errors. First, they 
used an incorrect scaling of the horizon mass during the radiation era, which 
should read
\begin{equation}
\sigma_{{\rm hor}} = \sigma_{{\rm hor}}(M_{{\rm eq}}) \left( 
	\frac{M}{M_{{\rm eq}}} \right)^{(1-n)/4} \,.
\end{equation}
Secondly, their COBE normalization was incorrect (too low) by a factor of 
around twenty. With these corrections, the constraint on $n$ tightens 
consider\-ably,\cite{gl} to become $n \lesssim 1.25$, rather than the $1.4$ 
to $1.5$ they quoted.

\section{With thermal inflation}

\begin{figure}[t]
\centering 
\leavevmode\epsfysize=7.7cm \epsfbox{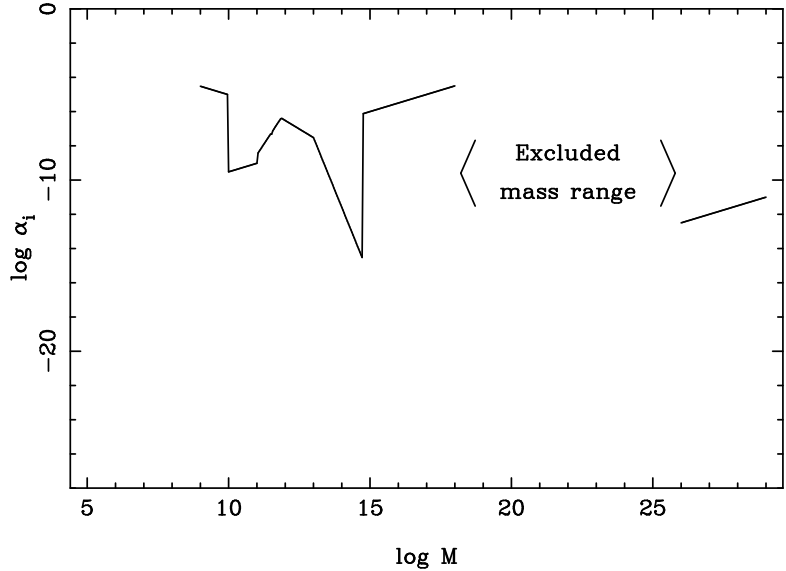}\\ 
\caption[fig2]{\label{fig2} PBH constraints modified to include thermal 
inflation.} 
\end{figure}

We model thermal inflation as occurring from $T = 10^7$ GeV down to the 
supersymmetry scale $T = 
10^3$ GeV, then reheating back up to $10^7$ GeV. As we have seen, most of the 
interesting mass region contains PBHs forming before $T = 10^7$ 
GeV, which implies that the constraints are affected by thermal inflation. 
Three effects are important:
\vspace{-3pt}
\begin{itemize}
\setlength{\itemsep}{-3pt}
\item Dilution of black holes during thermal inflation.
\item A change in the correspondence of scales: COBE scales 
leave the horizon closer to the end of inflation.
\item A mass range which enters the horizon before thermal 
inflation, but leaves again during it. No new perturbations are generated on 
this scale during thermal inflation, so from the horizon mass formula we find 
a missing mass range between $10^{18}$g and $10^{26}$g in which black holes 
won't form. Thermal inflation at higher energy could exclude masses below 
$10^{15}$g.
\end{itemize}
The dilution effect is shown in Fig.~\ref{fig2}; typical constraints now lie 
around $10^{-10}$ rather than $10^{-20}$. Taking all the effects into 
account,\cite{gl} the constraint on the spectral index weakens to $n \lesssim 
1.3$.

\begin{figure}[t]
\centering 
\leavevmode\epsfysize=7.7cm \epsfbox{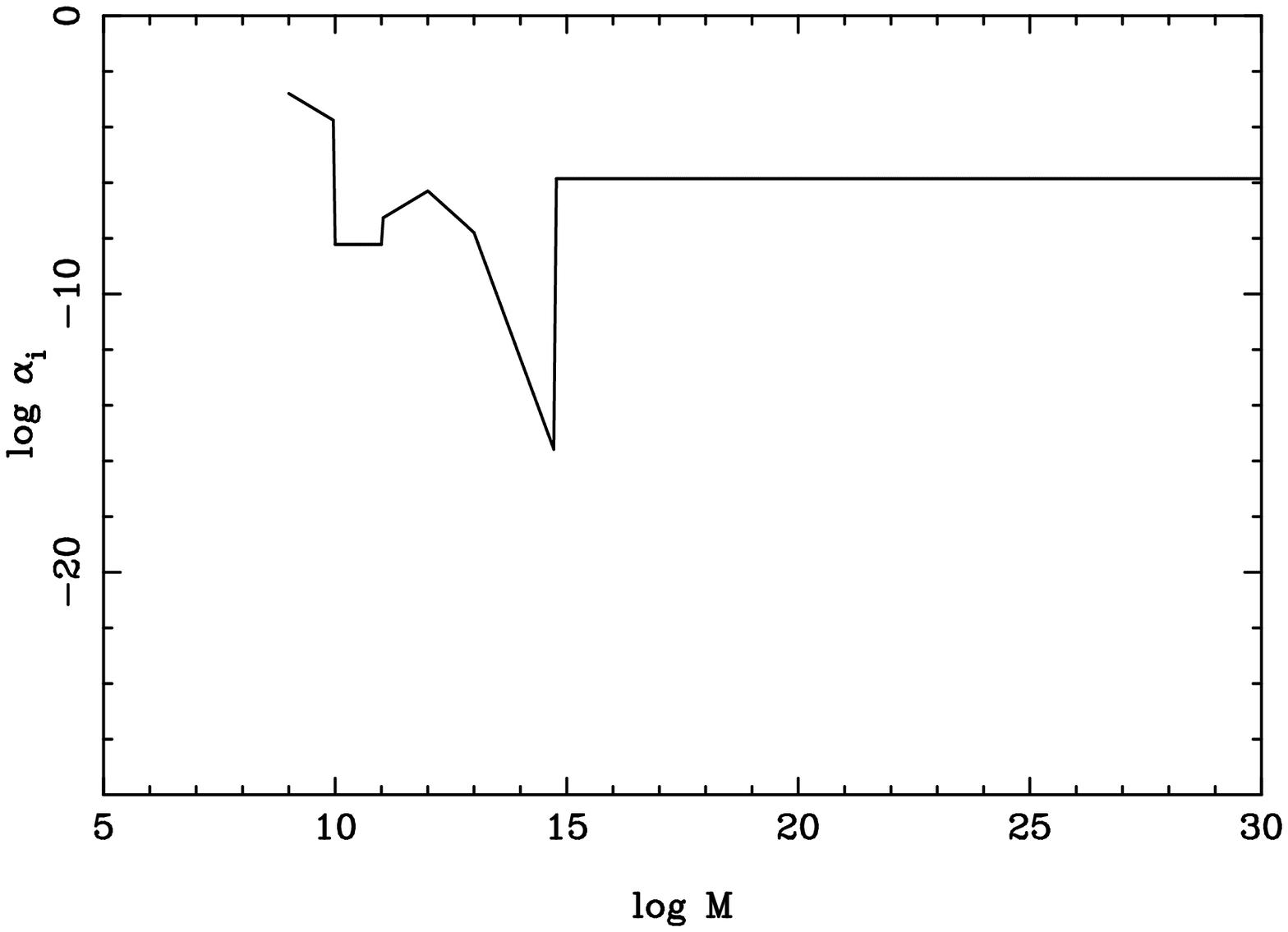}\\ 
\caption[fig3]{\label{fig3} PBH constraints modified for prolonged moduli 
domination.} 
\end{figure} 

\section{Cosmologies with moduli domination}

Another alternative is a prolonged early period of matter 
domination.\cite{glr} For 
example, moduli fields may dominate, and in certain parameter regimes can 
decay before nucleosynthesis. Various assumptions are possible; here we'll 
assume moduli domination as soon as they start to oscillate (around $10^{11}$ 
GeV). Part of the interesting range of PBH masses forms during moduli 
domination rather than radiation domination. Fig.~\ref{fig3} shows the 
constraints in this case, and with moduli domination the limit on $n$ again 
weakens\cite{glr} to $n \lesssim 1.3$.

\section{Conclusions}

Although PBH constraints are an established part of modern cosmology, they 
are sensitive to the entire cosmological evolution. In the standard 
cosmology, a power-law spectrum is constrained to $n < 1.25$, presently the 
strongest observational constraint on $n$ from any source. Alternative 
cosmological histories can weaken this to $n < 1.30$, and worst-case 
non-gaussianity \cite{BP} can weaken this by another 0.05 or so, though 
hybrid models giving constant $n$ give gaussian perturbations. Finally, we 
note that while the impact of the cosmological history on the 
density perturbation constraint is quite modest due to the exponential 
formation rate, the change can be much more significant for other formation 
mechanisms.

\section*{Acknowledgments}
ARL is supported by the Royal Society and AMG by PPARC. We thank Toni Riotto 
for collaboration on the moduli-dominated cosmology, and Bernard Carr and Jim 
Lidsey for discussions.


\end{document}